\documentclass{article}

\usepackage{PRIMEarxiv}

\usepackage[utf8]{inputenc} 
\usepackage[T1]{fontenc}    
\usepackage{hyperref}       
\usepackage{url}            
\usepackage{booktabs}       
\usepackage{amsfonts}       
\usepackage{nicefrac}       
\usepackage{microtype}      
\usepackage{lipsum}
\usepackage{amsmath} 
\usepackage{amssymb}
\usepackage{algorithm}
\usepackage{algpseudocode}
\usepackage{fancyhdr}       
\usepackage{graphicx}       
\usepackage{natbib}
\usepackage{float}
\usepackage{subfigure}
\usepackage{subcaption}
\usepackage{todonotes}
\usepackage{enumitem}

\graphicspath{{media/}}     

\title{DiffVolume: Diffusion Models for Volume Generation in Limit Order Books
}

\author{
  Zhuohan Wang \\
   Department of Informatics \\
  King's College London \\
  London, United Kingdom\\
  \texttt{zhuohan.wang@kcl.ac.uk}
  \And 
  Carmine Ventre \\
   Department of Informatics \\
  King's College London \\
  London, United Kingdom\\
  \texttt{carmine.ventre@kcl.ac.uk}
}

\begin{document}
\maketitle

\begin{abstract}
    Modeling limit order books (LOBs) dynamics is a fundamental problem in market microstructure research. In particular, generating high-dimensional volume snapshots with strong temporal and liquidity-dependent patterns remains a challenging task, despite recent work exploring the application of Generative Adversarial Networks to LOBs. In this work, we propose a conditional \textbf{Diff}usion model for the generation of future LOB \textbf{Volume} snapshots (\textbf{DiffVolume}). We evaluate our model across three axes: (1) \textit{Realism}, where we show that DiffVolume, conditioned on past volume history and time of day, better reproduces statistical properties such as marginal distribution, spatial correlation, and autocorrelation decay; (2) \textit{Counterfactual generation}, allowing for controllable generation under hypothetical liquidity scenarios by additionally conditioning on a target future liquidity profile; and (3) \textit{Downstream prediction}, where we show that the synthetic counterfactual data from our model improves the performance of future liquidity forecasting models. Together, these results suggest that DiffVolume provides a powerful and flexible framework for realistic and controllable LOB volume generation.
\end{abstract}


\section{Introduction}

Limit order books (LOBs) are the backbone of modern financial markets, capturing real-time information on pending buy and sell orders at various price levels \cite{gould2013limit}. Among the core components of LOBs, order volume at different price levels serves as a crucial indicator of market liquidity, price discovery, and trading opportunities. Accurate modeling of LOB volumes enables practitioners and researchers to improve trading strategies, optimize market making, and conduct robust market impact analyses \cite{bouchaud2018trades}. Therefore, the task of volume generation is fundamental for both theoretical insights and practical financial applications.

Most prior research focuses primarily on event-level generation, with the objective of simultaneously generating price and volume information, which often leads to less accurate and realistic results~\cite{li2020generating, coletta2021towards, nagy2023generative}. In particular, recent work by \citet{cont2023limit} applies Generative Adversarial Networks (GANs) to LOB modeling but only considers six price levels of a single big tick stock. Moreover, GAN-based models are prone to training instability and mode collapse, which can significantly impair their ability to replicate complex volume patterns across time and depth. Therefore, there remains a clear need for advanced modeling techniques that address these limitations in a comprehensive way.

To address these limitations, we introduce \textbf{DiffVolume}, a diffusion model-based framework specifically designed for the generation of future LOB volume snapshots. The proposed framework offers three main contributions:
\begin{itemize}[left=10pt]
    \item[${(1)}$] We develop a novel diffusion model architecture conditioning on past volume trajectories and time of day, which captures the spatial correlation structures and intricate temporal dependencies presented in LOB volume data;
    
    \item[${(2)}$] By additionally conditioning on a target future liquidity profile, we implement counterfactual generation under hypothetical liquidity scenarios;
    
    \item[${(3)}$] We demonstrate the quality and practical utility of the generated data by (i) evaluating the distribution of its data points vis-a-vis the original dataset; and (ii) showing how a downstream liquidity prediction task can be solved more effectively.
\end{itemize}

\section{Related Work}

We will discuss related research in three parts. 

\smallskip 
\noindent \textbf{Limit Order Book Dynamics.} The dynamics of LOBs have been extensively studied through various modeling approaches. Poisson processes models treat order arrivals and cancellations as independent events, characterized by a constant or time-varying intensity, providing a simplified representation of order flows \cite{bauwens2009modelling}. Hawkes processes models extend this framework by incorporating self-exciting properties, effectively modeling the clustering of order events, and capturing the feedback loops inherent in market activity \cite{bacry2015hawkes, filimonov2012quantifying}. Agent-based models simulate the interactions of heterogeneous traders with distinct strategies, allowing a detailed analysis of market microstructure phenomena and emergent behaviors such as volatility clustering and market crashes \cite{paddrik2012agent, byrd2019abides, amrouni2021abides, balch2019evaluate}.

\smallskip 
\noindent \textbf{Generative Modeling of Limit Order Books.} Generative modeling of LOBs can be primarily classified into two categories: autoregressive models and GANs. Autoregressive models generate LOB data sequentially, leveraging previous historical data points to predict future order book states. \citet{hultin2023generative} propose a model based on recurrent neural networks (RNNs) that decomposes the joint distribution of LOB transitions into conditional probabilities over order type, price, size, and time delay, each modeled with a dedicated RNN. Similarly, \citet{nagy2023generative} present an end-to-end autoregressive model using structured state-space models, which tokenizes message streams and generates realistic LOB transitions. GAN-based models, on the other hand, employ adversarial training to generate realistic synthetic limit order books. \citet{li2020generating} introduce Stock-GAN, a conditional Wasserstein GAN designed to generate realistic stock market order streams by capturing historical dependence and mimicking auction mechanisms. \citet{coletta2021towards} propose a Conditional GAN framework that reacts to current market states and allows agent interaction within a simulation, showing enhanced realism and responsiveness. Meanwhile, \citet{cont2023limit} develop a GAN framework that arguably learns the conditional distribution of future LOB volume states, capturing both stylized facts and some market impact patterns. However, their model can only be applied on large tick stock and limited price levels that are often asymmetric between the two sides of the book. DiffVolume accurately models the volume dynamics across many price levels for both bids and asks, no matter the tick size. 

\smallskip 
\noindent \textbf{Diffusion Models Theory.}  
The foundational idea of diffusion models is inspired by thermodynamic diffusion \cite{sohl2015deep}, which proposes to train generative models by adding Gaussian noise over multiple steps and learning to reverse the process. \citet{ho2020denoising} are the first to propose the Denoising Diffusion Probabilistic Model. 
Their method involves adding noise in a forward process and denoising in a reverse process to generate high-quality images. Subsequent work focuses on improving sampling quality and training efficiency ~\cite{nichol2021improved, song2020denoising}. Diffusion models are also closely related to score-based generative modeling, where the model learns the gradient of the data density through denoising score matching ~\cite{vincent2011connection, song2019generative}. This connection has been formally unified under the stochastic differential equation framework by \cite{song2020score}. Recent advances further clarify and optimize the design space of diffusion models under this unified view. For example, \citet{karras2022elucidating} analyze the score matching and noise schedules in depth, leading to improvements in sample quality and generation speed. This line of work provides the theoretical foundation for applying diffusion models to structured data domains such as financial data \cite{koa2023diffusion, wang2024financial}.

\section{Methodology}\label{section:theory}
In this section, we will briefly summarize conditional denoising diffusion probabilistic models from the score-based  perspective and describe the proposed DiffVolume architecture in detail.

\subsection{Generative Diffusion Models}

\smallskip 
\noindent \textbf{Denoising Diffusion Probabilistic Models.} Denoising Diffusion Probabilistic Models (DDPMs)~\cite{sohl2015deep, ho2020denoising, luo2022understanding, chan2024tutorial} consist of two complementary stochastic processes, a forward (noising) process and a reverse (generative) process. In the forward process, Gaussian noise is progressively added to the data through a fixed Markov chain, transforming it into a nearly pure noise distribution. The reverse process learns to invert this transformation by iteratively denoising the corrupted data, thereby recovering samples from the data distribution via learned transitions.

Consider a noise strength sequence $0<\beta_1< \beta_2 <\cdots< \beta_N <1$ and forward noising process $p(\mathbf{x}_i|\mathbf{x}_{i-1})=\mathcal{N}(\mathbf{x}_i|\sqrt{1-\beta_i}\mathbf{x}_{i-1}, \beta_i\mathbf{I})$. With $\alpha_i:=\prod_{j=1}^i(1-\beta_j)$, we have $p_{\alpha_i}(\mathbf{x}_i|\mathbf{x}_{0})=\mathcal{N}(\mathbf{x}_i|\sqrt{\alpha_i}\mathbf{x}_{0}, (1-\alpha_i)\mathbf{I})$, and the perturbed data distribution is given by $p_{\alpha_i}(\tilde{\mathbf{x}})=\int p_{\mathrm{data}}(\mathbf{x}) p_{\alpha_i}(\tilde{\mathbf{x}}|\mathbf{x})d\mathbf{x}$. The noise scale is predetermined to approximately satisfy $p_{\alpha_N}(\tilde{\mathbf{x}}) \sim \mathcal{N}(\tilde{\mathbf{x}}|\mathbf{0}, \mathbf{I})$, i.e., the terminal distribution is nearly independent of $p_{\mathrm{data}}$. The reverse denoising process can be written as $p_{\theta}(\mathbf{x}_{i-1}|\mathbf{x}_i)=\mathcal{N}(\mathbf{x}_{i-1}|\frac{1}{\sqrt{1-\beta_i}}(\mathbf{x}_i+\beta_i \mathbf{s}_{\theta}(\mathbf{x}_i,i)),\beta_i\mathbf{I})$, where $\mathbf{s}_{\theta}(\mathbf{x}_i,i)$ is called the \textit{score}\cite{song2019generative}. The training objective is a sum of local denoising score-matching objectives, i.e., finding $\theta^*$ that minimizes
\begin{equation}
    \sum_{i=1}^N (1-\alpha_i) \mathbb{E}_{p_{\mathrm{data}}(\mathbf{x})}\mathbb{E}_{p_{\alpha_i}(\tilde{\mathbf{x}}|\mathbf{x})}\big[||\mathbf{s}_{\theta}(\tilde{\mathbf{x}},i)-\nabla_{\tilde{\mathbf{x}}}\log p_{\alpha_i}(\tilde{\mathbf{x}}|\mathbf{x})||^2_2 \big].
\label{eq:ddpm-training}
\end{equation}
 
\smallskip 
\noindent \textbf{DDPMs and Stochastic Differential Equations.} \citet{song2020score} demonstrate that DDPMs can be understood from the perspective of stochastic differential equations (SDEs). Let $\{\mathbf{x}(t) \}_{t=0}^T$ be a stochastic diffusion process indexed by a continuous time variable $t\in[0, T]$, evolving from $\mathbf{x}(0) \sim p_0$, the true data distribution, to $\mathbf{x}(T) \sim p_T$, approximately the tractable prior distribution. Denote the probability density function of $\mathbf{x}(t)$ by $p_t(\mathbf{x})$ and the transition kernel from $\mathbf{x}(s)$ to $\mathbf{x}(t)$ by $p_{st}(\mathbf{x}(t)|\mathbf{x}(s))$, for $0\leq s<t \leq T$. Then, we can use an SDE to represent such a forward diffusion process: 
\begin{equation}
    d\mathbf{x}=\mathbf{f}(\mathbf{x},t)\ dt + g(t) \ d \mathbf{w},
\label{eq:sde-forward-process}
\end{equation}
where $\mathbf{f}(\mathbf{x},t) dt$ is referred to as the \textit{drift} term, and $ g(t) d \mathbf{w}$ is referred to as the \textit{diffusion} term. Here, $\mathbf{w}$ is a standard Wiener process and $d\mathbf{w} \sim \mathcal{N}(0, dt \mathbf{I})$. The synthetic data generation process is the reverse process of Eq. (\ref{eq:sde-forward-process}), which is also an SDE~\cite{anderson1982reverse}:
\begin{equation}
    d\mathbf{x}=[\mathbf{f}(\mathbf{x},t)-g^2(t) \nabla_{\mathbf{x}}\log p_t(\mathbf{x})]\ dt +  g(t)\ d \bar{\mathbf{w}},
\label{eq:sde-backward-process}
\end{equation}
where $\bar{\mathbf{w}}$ is a reverse-time Wiener process and  $\nabla_{\mathbf{x}}\log p_t(\mathbf{x})$ is the score of the marginal distribution corresponding to each $t$. It starts from an initial noise sample $\mathbf{x}(T) \sim p_T$ and gradually denoises it step by step following Eq. (\ref{eq:sde-backward-process}). Theoretically, if $T \rightarrow \infty$, we obtain $\mathbf{x}(0) \sim p_0$. To estimate $\nabla_{\mathbf{x}}\log p_t(\mathbf{x})$, the score network $\mathbf{s}_{\mathbf{\theta}}(\mathbf{x}, t)$ is trained using the objective function 
\begin{equation}
    \!\!\kappa(t) \mathbb{E}_t  \mathbb{E}_{\mathbf{x}(0)} \mathbb{E}_{\mathbf{x}(t)|\mathbf{x}(0)}\!\big[||\mathbf{s}_{\theta}(\mathbf{x}(t),t)\!- \!\nabla_{\mathbf{x}(t)}\!\log p_{0t}(\mathbf{x}(t)|\mathbf{x}(0))||^2_2 \big],
\label{eq:sde-training}
\end{equation}
where $\kappa:[0,T] \rightarrow \mathbb{R}^+$ is a positive weight and $t \sim \mathcal{U}[0,T]$. Eq. \eqref{eq:sde-training} is a continuous generalization of Eq. \eqref{eq:ddpm-training}. Typically, the continuous form of the DDPM forward process is chosen to be
\begin{equation}
    d\mathbf{x}=-\frac{\beta(t)}{2} \mathbf{x}\  dt + \sqrt{\beta(t)} \ d\mathbf{w},
\label{ddpm-continuous-forward-process}
\end{equation}
i.e., $\mathbf{f}(\mathbf{x},t)=-\frac{\beta(t)}{2} \mathbf{x}$ and $g(t) = \sqrt{\beta(t)}$. Substituting $\mathbf{f}(\mathbf{x},t)$ and $g(t)$ in (\ref{eq:sde-backward-process}), we can get the backward process in SDE form for DDPM. 

\smallskip 
\noindent \textbf{Conditional DDPMs.} How do we inject the conditioning $\mathbf{c}$ into the training and sampling process? Here we follow the  \textit{classifier-free guidance} approach~\cite{ho2022classifier}, combining the conditional and unconditional models as follows:
\begin{equation}
    \nabla_{\mathbf{x
    }} \log \tilde{p}(\mathbf{x}|\mathbf{c})=\omega\nabla_{\mathbf{x}} \log p(\mathbf{x}|\mathbf{c})+(1-\omega)\nabla_{\mathbf{x}} \log p(\mathbf{x}),
\label{eq:classifier-free-guidance}
\end{equation}
where $\nabla_{\mathbf{x}} \log p(\mathbf{x}|\mathbf{c})$ represents the conditional  and $\nabla_{\mathbf{x}} \log p(\mathbf{x})$ represents the unconditional score, corresponding to the conditional and unconditional model distributions. Eq. (\ref{eq:classifier-free-guidance}) reduces to the unconditional score when $\omega=0$, or recovers the conditional score  when $\omega=1$. 

\subsection{DiffVolume Architecture}
\begin{figure*}[t]  
    \centering  
    \includegraphics[width=\textwidth]{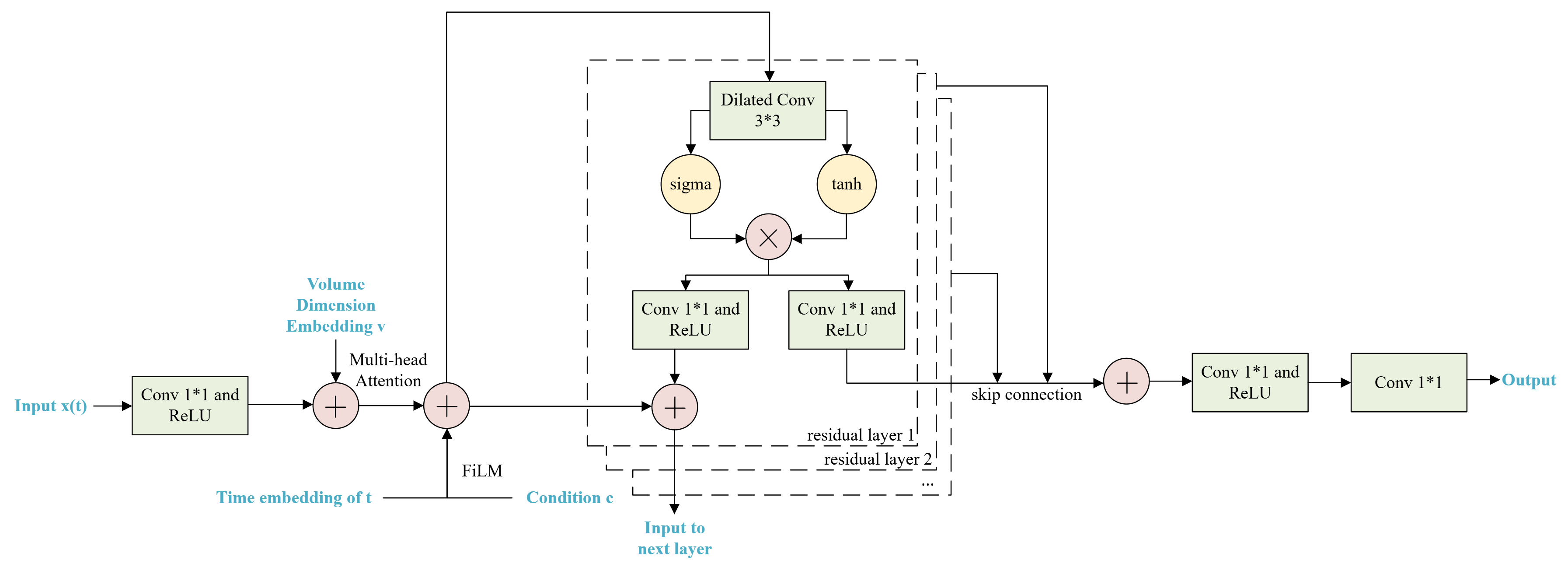}  
    \caption{Illustration of DiffVolume Architecture.}
    \label{fig:architecture}  
\end{figure*}
DiffVolume is a conditional score-based diffusion model designed to estimate the score function $\mathbf{s}_{\theta}(\mathbf{x}(t),t,\mathbf{c})$. The architecture is shown in Figure \ref{fig:architecture}. At each diffusion step $t$, the model takes as input a noised volume snapshot denoted by $\mathbf{x}(t)$. The architecture incorporates two embeddings \textit{price level embeddings} $ \mathbf{v}$, \textit{time step embeddings} $t$, and an external \textit{conditioning context} $\mathbf{c}$, which includes variables such as past volume trajectories, time of day and target future liquidity (if applicable). Details on the construction of $\mathbf{v}$ and $t$ are provided below, while the structure of $\mathbf{c}$ is described in Subsection~\ref{subsection: training and sampling}.

The input $\mathbf{x}(t)$ is first processed by a $1\times 1$ convolutional layer followed by a ReLU activation. This projects the corrupted volume vector into a higher-dimensional latent feature space, making it more suitable for capturing complex interactions. The resulting latent feature is then augmented with learnable \textit{price level embeddings} $ \mathbf{v}$, where each discrete price level index is mapped to a continuous dense vector. These embeddings enable the model to distinguish between spatially different price levels in the order book. To capture the dependencies across price levels, the combined input is passed through a multi-head self-attention module~\cite{vaswani2017attention}.

To encode the diffusion step $t$, we adopt a sinusoidal embedding scheme followed by a two-layer MLP with SiLU activations. To incorporate the conditioning information from both $t$ and $\mathbf{c}$, we adopt the Feature-wise Linear Modulation (FiLM) mechanism~\cite{perez2018film}. Specifically, the FiLM layers learn affine transformations (scaling and shifting) applied to the intermediate features, parameterized by functions of $t$ and $\mathbf{c}$. This allows the model to adjust its internal computations dynamically, enabling more flexible and adaptive generation.

The core of the DiffVolume architecture is a stack of 32 residual convolutional layers, inspired by WaveNet-style dilated convolutions \cite{van2016wavenet}. Each residual block begins with a dilated $3 \times 3$ convolutional layer, which increases the receptive field without increasing model depth. The output is split into two branches: one activated with a \texttt{tanh} function, and the other with a \texttt{sigmoid} function. The two activations are combined through element-wise multiplication, forming a \textit{soft gating mechanism} that allows the network to selectively pass or suppress signals.

The gated output is then passed through two $1 \times 1$ convolutional layers, each followed by a ReLU activation. A residual connection is added between the input and output of each block, facilitating gradient flow and stability during training. In addition, each residual layer outputs a skip connection, which is accumulated across all layers and routed to the final output head. The aggregated skip connections are combined and passed through a final output sequence: a $1 \times 1$ convolution followed by ReLU activation, and another $1 \times 1$ convolution to produce the final prediction. 

This novel architecture enables the model to integrate fine-grained spatial dependencies within each price level and broader contextual signals across the book and conditioning inputs, allowing it to accurately model the high-dimensional structure of LOB volume trajectories.

\section{Experimental Setup}\label{section:methodology}
In this section, we provide details on the dataset used in our experiments, the preprocessing steps applied to the raw data, as well as the training and sampling procedures.

\subsection{Data}
We use the LOBSTER data\footnote{https://lobsterdata.com/} as our LOB data source \cite{huang2011lobster}. The platform provides Level-3 data, allowing for a full-fidelity reconstruction of the LOB dynamics. For each day, the dataset contains a message file, which consists of orders sent to an exchange (primarily market, limit, and (partial) cancellation orders), and a snapshot file, which includes the corresponding order book states. To construct our dataset, we sample one snapshot per second from the snapshot files. Each snapshot includes the top 10 levels on both the bid and ask sides, resulting in a volume representation of shape $[20, 1]$ per time step.

In particular, we train, evaluate, and test our model separately on four stocks, which are MU (Micron, big tick stock), AAPL (Apple, medium tick stock), ADBE (Adobe, small tick stock) and ZM (Zoom, small tick stock). Stocks are classified as big, medium, or small tick size according to how their tick size compares to the typical bid-ask spread. Variations in tick size can affect bid-ask spreads, order book depth, and execution costs, thus shaping the strategies of market participants and the dynamics of price formation \cite{briola2025hlob}. For each stock, we use 16 consecutive trading days for training (1 February to 23 February 2023), 1 day for validation (24 February  2023), and the final 2 days for testing (27-28 February 2023).

\subsection{Preprocessing}

To mitigate the non-stationary behavior typically observed in limit order books—particularly around market open and close—we exclude the first and last 30 minutes of each continuous trading session from our analysis. As a result, we retain only snapshots from the 5.5-hour interval between 10:00 AM and 3:30 PM each day. To ensure data quality and prevent computational inconsistencies, we remove all snapshots with missing values. Given that such instances are extremely rare, their exclusion has a negligible effect on the overall statistical properties of the dataset. Setting the sampling frequency to 1 second results in $\sim 310$k order book snapshots for training, $\sim 20$k snapshots for validation, and $\sim 40$k snapshots for testing. 

To mitigate the impact of extreme outliers in volume data, we apply a clipping procedure that caps volume values at the 99th percentile for each price level. This enhances model stability and ensures that learning is not dominated by rare but disproportionately large volume observations. We also adopt a square-root-based normalization scheme as shown in Eq. (\ref{eq:preprocess_normalization}). It is coupled with a scaling constant, to stabilize the input distribution and make the training process more effective: 
\begin{equation}
    x_{normalized} = \frac{\sqrt{x}}{const},
    \label{eq:preprocess_normalization}
\end{equation}
where $const=15$ during real training.
\begin{figure*}[!htbp]  
    \centering  
    \includegraphics[width=\textwidth]{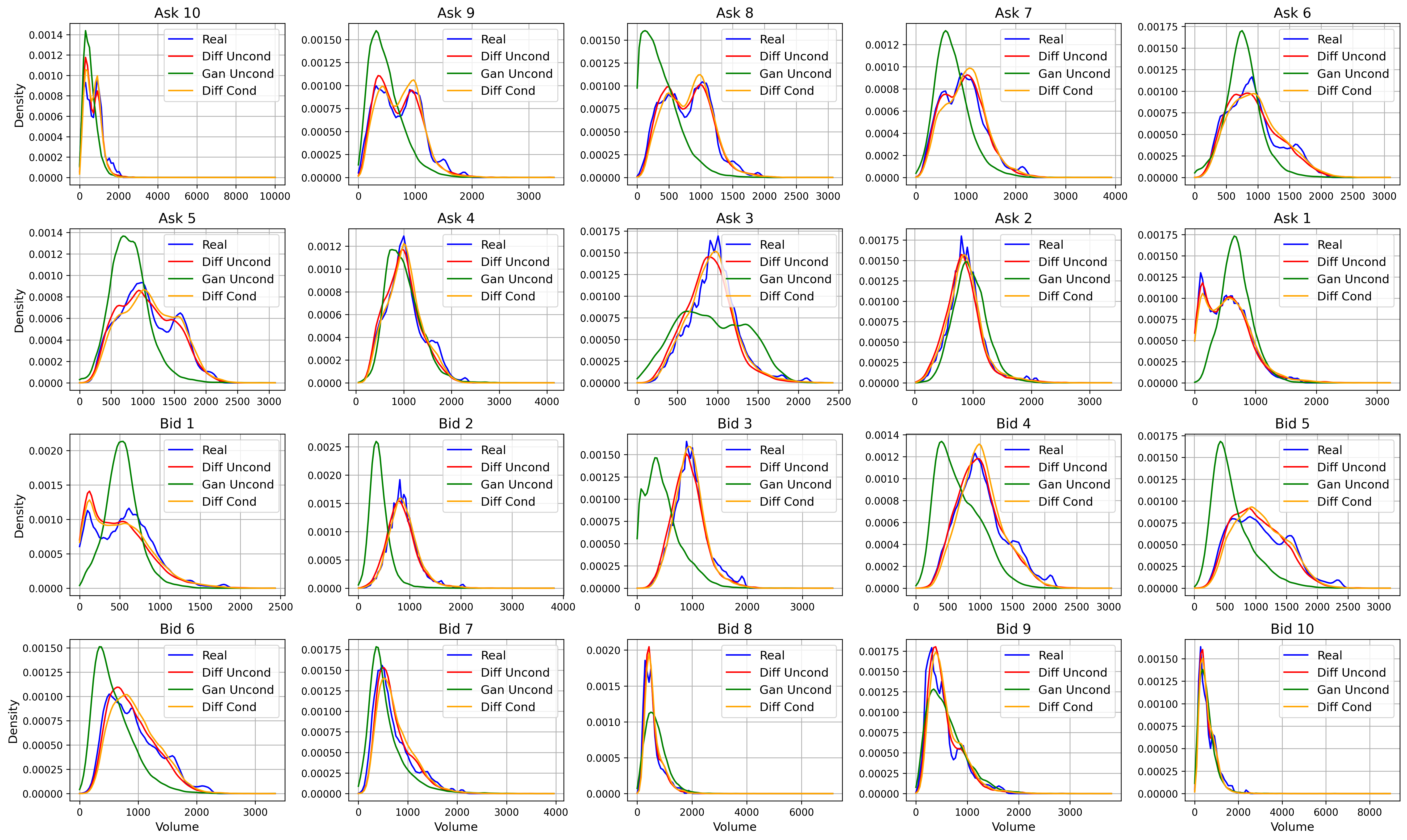}  
    \caption{Marginal Volume Distribution.}
    \label{fig:volume distribution}  
\end{figure*}
\subsection{Training and Sampling}\label{subsection: training and sampling}

We consider the task of modeling the conditional distribution of future LOB volume snapshots given multiple conditioning variables. Let $\mathbf{x}_{(t-L+1:t)} \in \mathbb{R}^{L \times D}$ denote the sequence of past volume snapshots over $L$ time steps, where each vector $\mathbf{x}_{(i)} \in \mathbb{R}^{D}$ represents volume across $D$ price levels (20 levels in our case) and $x_{(i)}^j$ represents the volume value at the $j$-th price level in time step $i$. Our objective is to generate a realistic sequence of future volumes $\mathbf{x}_{(t+1:t+L)} \in \mathbb{R}^{L \times D}$ based on a \textit{conditioning context} $\mathbf{c}$, which includes the past volume trajectory $\mathbf{x}_{(t-L+1:t)}$, the time of day $\mathbf{\tau}_{(t+1:t+L)}$ and, in counterfactual settings, the future liquidity indicator $\mathbf{\lambda}_{(t+1:t+L)}$. The length $L$ is set to $32$ in the training and sampling stage.

The time of day $\tau_{(t+1:t+L)}$ is constructed as a normalized scalar ratio reflecting the elapsed time since the market opening, calculated as
\begin{equation}
    \tau_{(i)} = \frac{\Delta t_i}{T_{total}}, \ \ \  for \ i \in \{t+1, ...t+L\},
    \label{eq:condition_t}
\end{equation}
where $\Delta t_i$ is the number of seconds elapsed from the start of the trading session, and $T_{total}=19800$ seconds denotes the total trading duration (5.5 hours). The resulting sequence $\mathbf{\tau}_{(t+1:t+L)} \in \mathbb{R}^{L \times 1}$ serves as a continuous temporal encoding aligned with each future snapshot. By conditioning on the time of the day, we provide temporal context that captures non-stationary behaviors over intraday horizons. 

The future liquidity indicator $\mathbf{\lambda}$ provides a coarse but informative signal reflecting the expected market depth over the prediction horizon. Each value in $\mathbf{\lambda}$ is calculated as the sum of the volume of the corresponding LOB snapshot:
\begin{equation}
    \lambda_{(i)} = \sum_{j=1}^{D} x_{(i)}^{j}, \ \ \  for \ i \in \{t+1, ...t+L\}.
    \label{eq:condition_liquidity}
\end{equation}
Conditioning the model on this sequence, we enable controlled generation, allowing the model to produce volume patterns that not only exhibit realistic microstructure dynamics but also align with target liquidity profiles. 


To effectively train the conditional diffusion model and prevent overfitting or instability, we adopt the early stopping \cite{goodfellow2016deep} and exponential moving average (EMA) mechanism of the network weights \cite{nichol2021improved}. Early stopping is used based on the validation loss, where training is terminated if the validation loss does not improve by at least 0.001 over a patience window of 100 epochs. EMA parameters are updated after each optimization step as $\theta_{EMA} \leftarrow \alpha \cdot \theta_{EMA} + (1-\alpha) \cdot \theta$, where $\alpha=0.999$. We train the model using Adam optimizer \cite{diederik2014adam} with a learning rate of $1 \times 10^{-4}$ and a batch size of $64$. We adopt the ancestral sampling procedure \cite{ho2020denoising}, a widely used method in DDPMs, to generate LOB volume snapshots.

\section{Experimental Results}
In this section, we address three key research questions. (1) To what extent do the generated samples exhibit statistical fidelity to real-world limit order book data? (2) Can DiffVolume generate meaningful counterfactual samples in response to varying liquidity conditions? (3) Are the counterfactually generated samples informative enough to support downstream applications, such as future liquidity prediction?

\subsection{Realism}\label{subsection: realism}
We begin by assessing the realism of the generated LOB volumes from different models, focusing on their ability to replicate key statistical properties observed in real market data. We compare four sources of volume data: (i) the empirical testing dataset (\textit{Real}), (ii) diffusion model without future liquidity condition (\textit{Diff Uncond}), (iii) WGAN-GP model without future liquidity condition (\textit{GAN Uncond}), and (iv) diffusion model with future liquidity condition (\textit{Diff Cond}). To ensure a fair comparison of generative quality, we primarily focus on \textit{GAN Uncond} and \textit{Diff Uncond}, where no strong future information is provided during training. The \textit{Diff Cond} model, although conditioned on future liquidity profiles, is included here for completeness, as it plays a key role in later sections and shows promise for future applications. We will take MU as an illustrative example for our plots and summarize the performance for the four stocks in Table~\ref{tab:realism}.

\smallskip \noindent \textbf{Marginal Volume Distributions.} Figure \ref{fig:volume distribution} presents kernel density estimates of marginal volume distributions at each price level. Real LOB volumes exhibit multi-modal and heavy-tailed characteristics. GAN-generated samples fail to reproduce these higher-order statistics, often yielding overly smooth unimodal distributions. Both \textit{Diff Uncond} and \textit{Diff Cond} accurately capture the skewness and kurtosis of empirical data, outperforming \textit{GAN Uncond}.

\begin{figure}[!htbp]  
    \centering  
    \includegraphics[width=\linewidth]{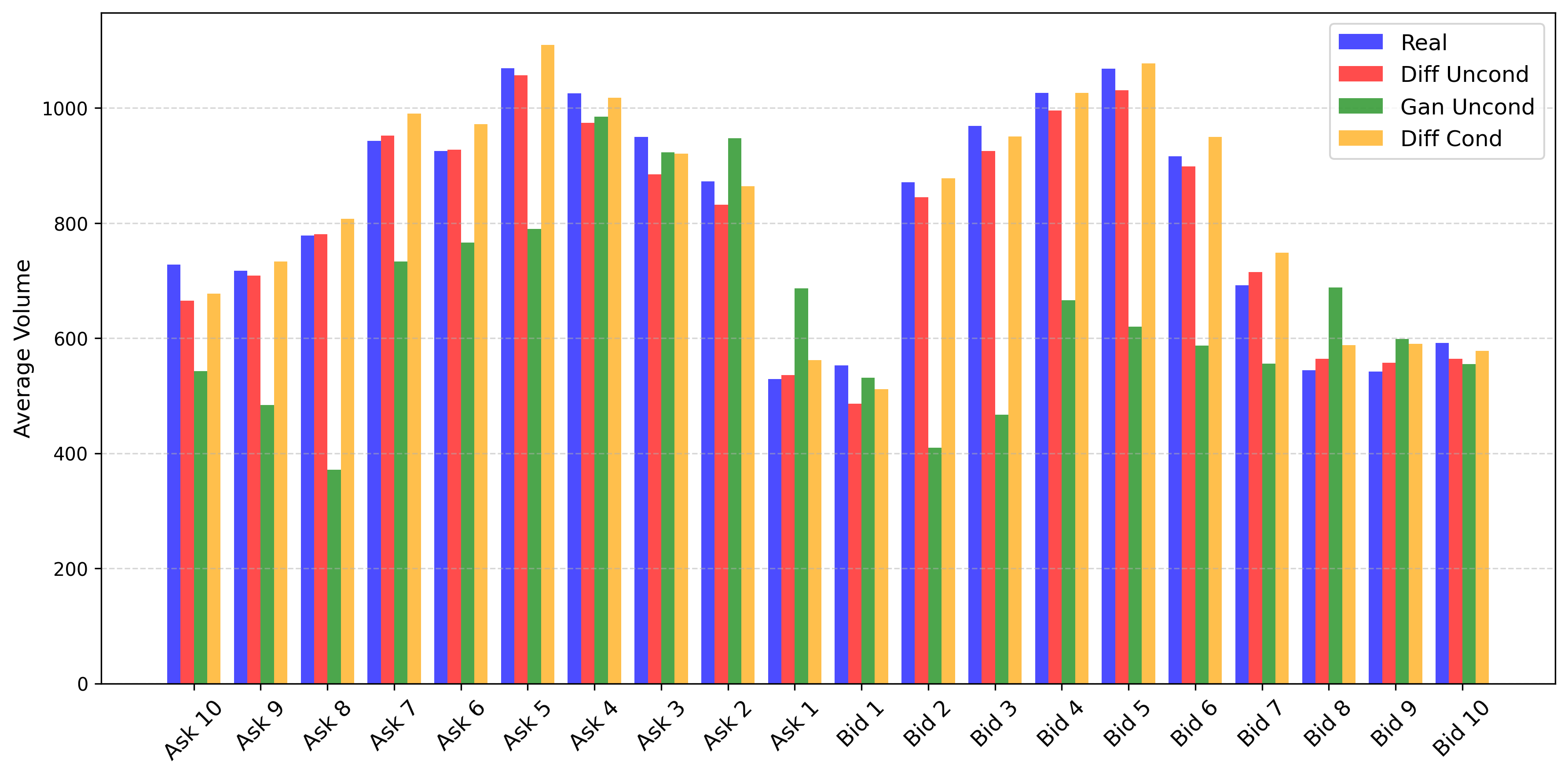}  
    \caption{Average Volume per Price Level.}
    \label{fig:volume average}  
\end{figure}

\smallskip \noindent \textbf{Average Volume per Price Level.} Figure \ref{fig:volume average} displays the average volume across each of the 10 ask and 10 bid price levels. The real volume exhibits a clear symmetric hump-shaped structure around the best bid and ask prices, with the highest concentration of liquidity at levels 3–6. \textit{GAN Uncond} underestimates volume at nearly all levels. In contrast, \textit{Diff Uncond} and \textit{Diff Cond} closely match the empirical averages, suggesting its ability to reproduce realistic liquidity profiles.

\begin{figure}[!htbp]  
    \centering  
    \includegraphics[width=0.8\linewidth]{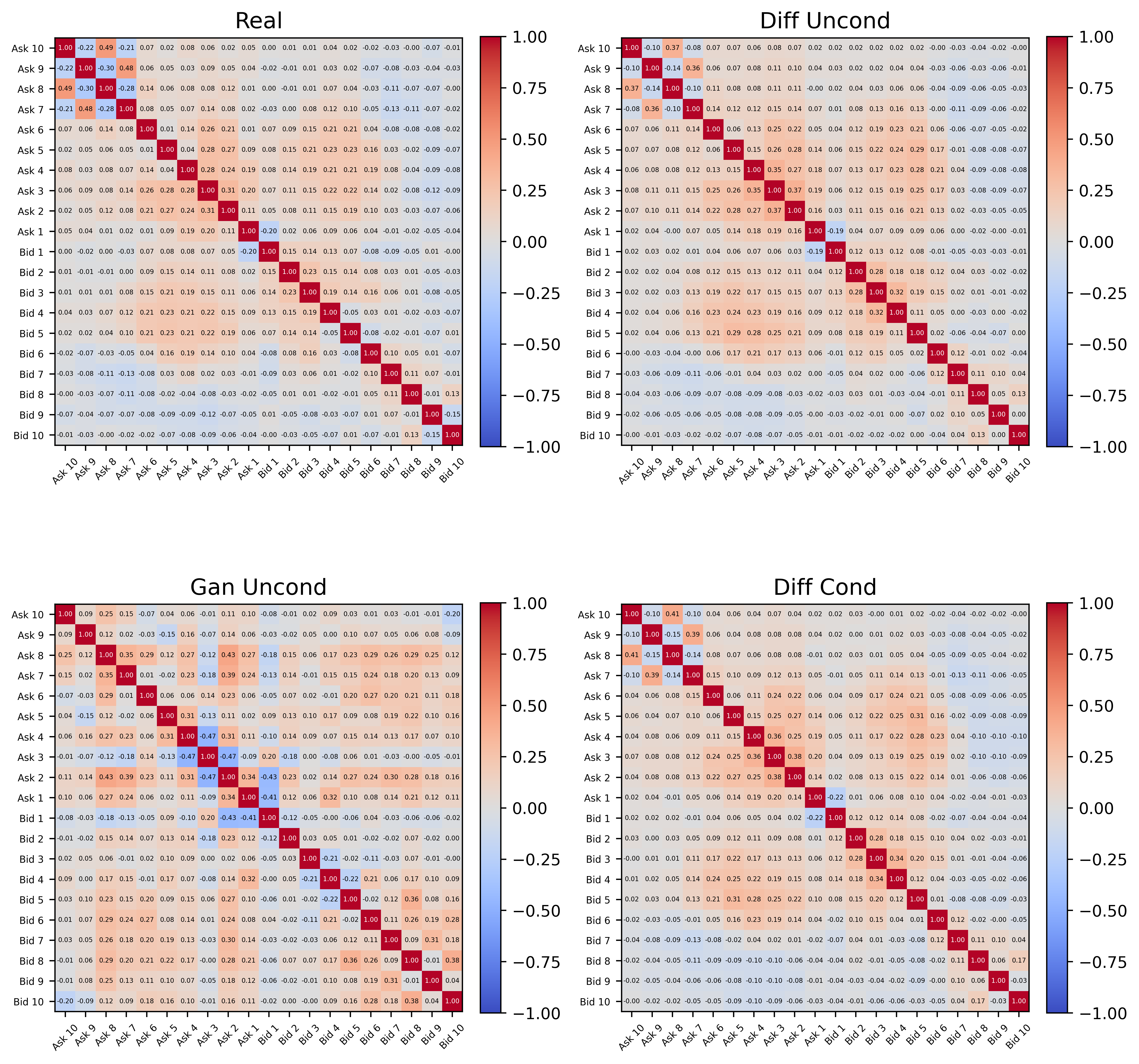}  
    \caption{Cross-sectional Volume Correlation.}
    \label{fig:volume correlation}  
\end{figure}

\begin{figure}[!htbp]  
    \centering  
    \includegraphics[width=0.8\linewidth]{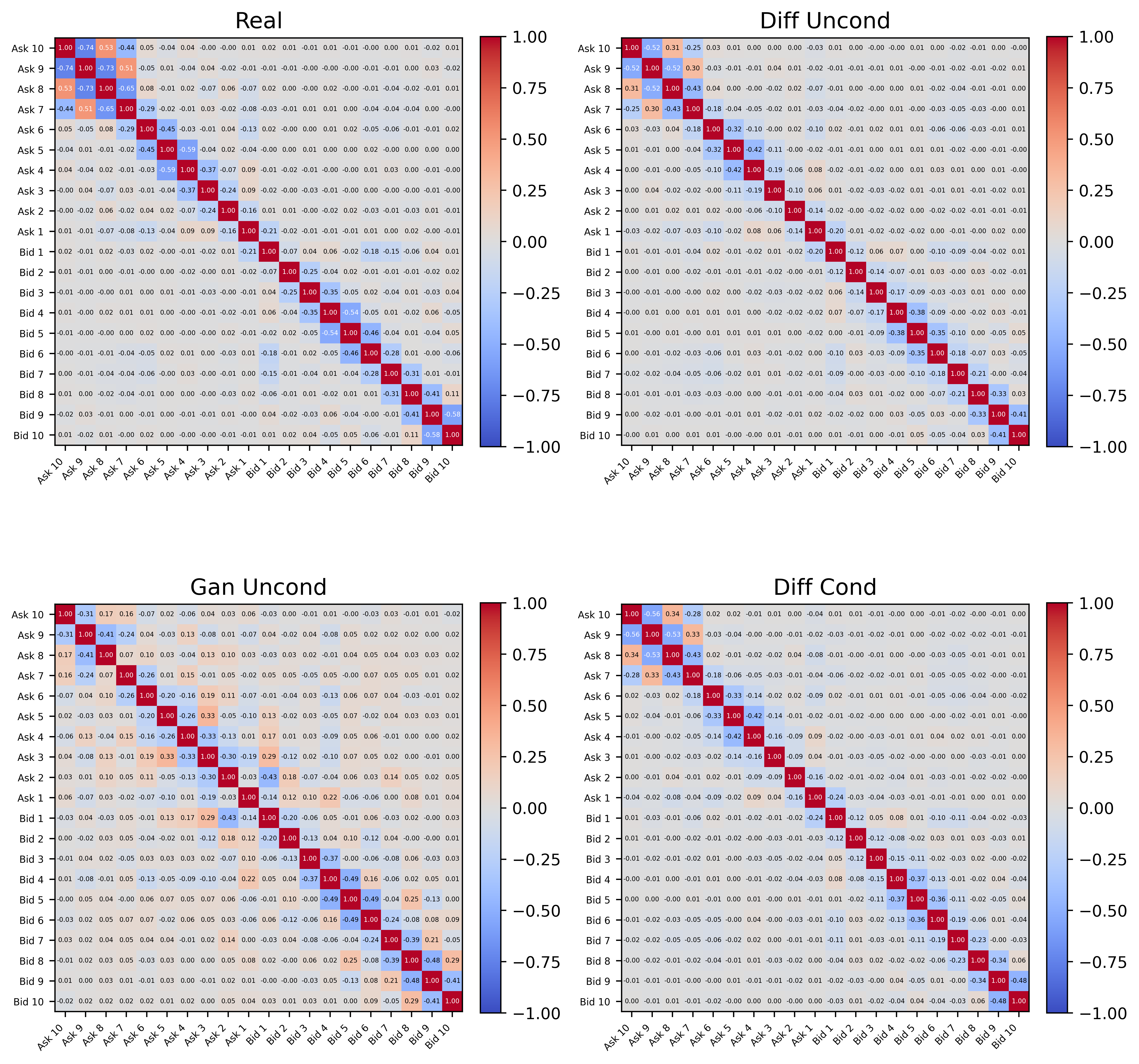}  
    \caption{Temporal Difference Volume Correlation.}
    \label{fig:volume temporal difference correlation}  
\end{figure}

\smallskip \noindent \textbf{Cross-sectional and Temporal Difference Correlation.} Figure \ref{fig:volume correlation} shows the cross-sectional volume correlation across 20 price levels (Ask 10 to Bid 10). Compared with \textit{GAN Uncond}, \textit{Diff Uncond} and \textit{Diff Cond} can better capture the spatial structure in the upper left quadrant (Ask 10 to Ask 7) and inter-side interactions (Ask 2 to Ask 6, Bid 2 to Bid 6). Figure \ref{fig:volume temporal difference correlation} is calculated from the first-order differences of the volume snapshots ($\Delta \mathbf{x}_{(i)} = \mathbf{x}_{(i)} - \mathbf{x}_{(i-1)}$). A notable characteristic in the temporal difference correlation structure is the negative correlation at adjacent price levels, which is captured by all three models. However, \textit{Diff Uncond} and \textit{Diff Cond} do better in the upper left quadrant (Ask 10 to Ask 7). 

\begin{figure*}[!htbp]  
    \centering  
    \includegraphics[width=\textwidth]{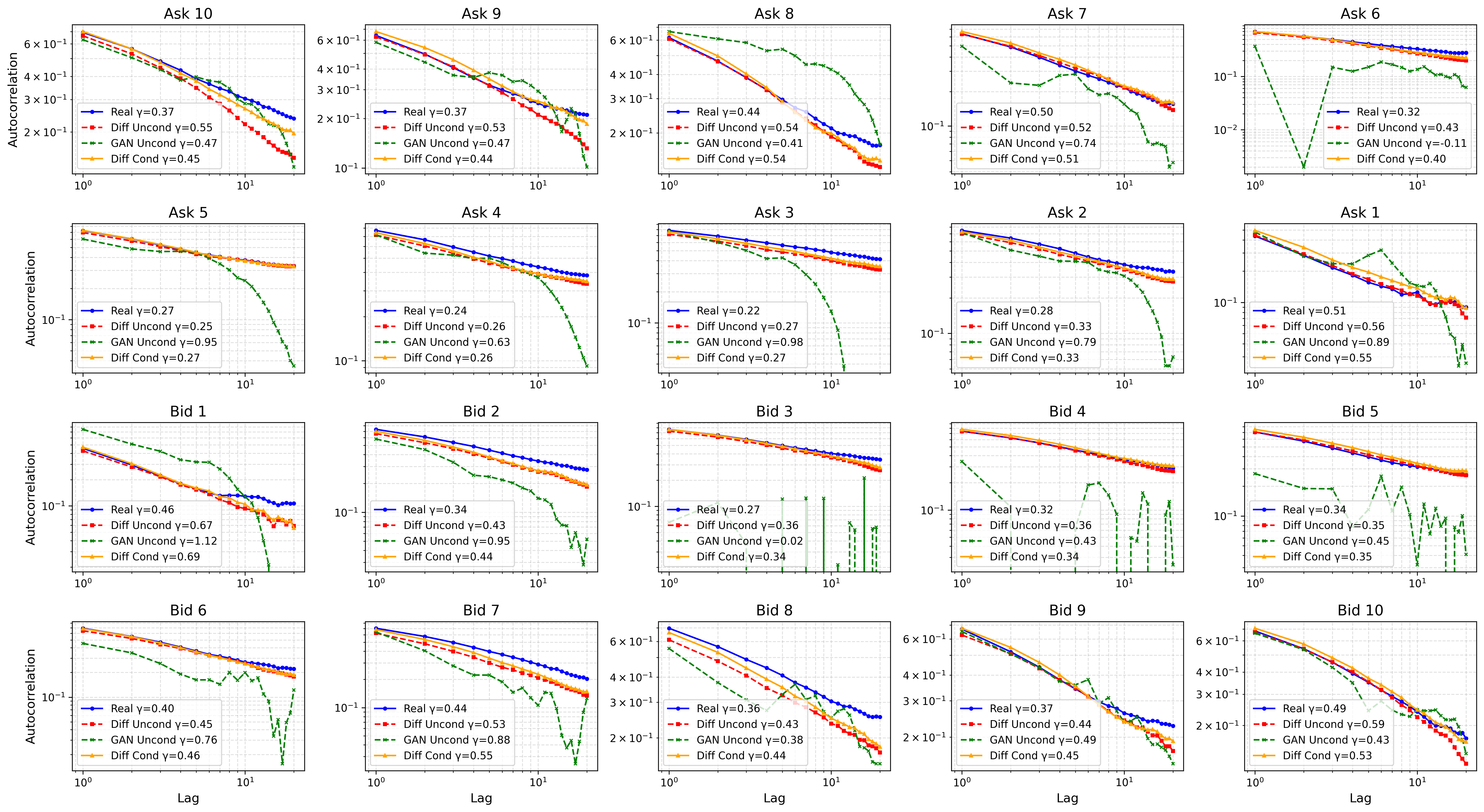}  
    \caption{Autocorrelation Decay.}
    \label{fig:autocorrelation decay}  
\end{figure*}

\smallskip \noindent \textbf{Autocorrelation Decay.} Finally, we evaluate the long memory properties of the generated volume series by analyzing autocorrelation decay at each price level, shown in Figure \ref{fig:autocorrelation decay} as a log-log plot. In empirical financial time series, volume autocorrelation is known to approximately follow a scaling law \cite{gould2013limit}:
\begin{equation}
    ACF(l) \sim C \cdot l^{-\gamma},
    \label{eq:ACF}
\end{equation}
where $ACF(l)$ denotes the lag-$l$ autocorrelation. The conditional diffusion model closely replicates these dynamics, while the unconditional diffusion model slightly underperforms with faster decay rates. In contrast, the GAN-based approach demonstrates unstable and unrealistic temporal correlations.
\begin{table*}\centering
\scalebox{0.77}{
\begin{tabular}{l|ccc|ccc|ccc|ccc|}
\cline{2-13}
                                       & \multicolumn{3}{c|}{MU}                                                                    & \multicolumn{3}{c|}{AAPL}                                                                    & \multicolumn{3}{c|}{ADBE}                                                                   & \multicolumn{3}{c|}{ZM}                                                                                                                   \\ \cline{2-13} 
                                       & \multicolumn{1}{c|}{Wasserstein}    & \multicolumn{1}{c|}{KL}             & KS             & \multicolumn{1}{c|}{Wasserstein}      & \multicolumn{1}{c|}{KL}             & KS             & \multicolumn{1}{c|}{Wasserstein}     & \multicolumn{1}{c|}{KL}             & KS             & \multicolumn{1}{c|}{Wasserstein}     & \multicolumn{1}{c|}{KL}                                    & KS                                    \\ \hline
\multicolumn{1}{|l|}{Diff Uncond}      & \multicolumn{1}{c|}{\textbf{40.34}} & \multicolumn{1}{c|}{\textbf{0.256}} & \textbf{0.091} & \multicolumn{1}{c|}{165.783}          & \multicolumn{1}{c|}{0.291}          & 0.25           & \multicolumn{1}{c|}{12.169}          & \multicolumn{1}{c|}{\textbf{0.344}} & \textbf{0.136} & \multicolumn{1}{c|}{24.509}          & \multicolumn{1}{c|}{0.199}                                 & 0.135                                 \\ \hline
\multicolumn{1}{|l|}{GAN Uncond}       & \multicolumn{1}{c|}{229.106}        & \multicolumn{1}{c|}{0.687}          & 0.31           & \multicolumn{1}{c|}{190.955}          & \multicolumn{1}{c|}{0.508}          & 0.258          & \multicolumn{1}{c|}{30.398}          & \multicolumn{1}{c|}{1.155}          & 0.348          & \multicolumn{1}{c|}{94.96}           & \multicolumn{1}{c|}{0.635}                                 & 0.335                                 \\ \hline
\multicolumn{1}{|l|}{Diff Cond}        & \multicolumn{1}{c|}{42.603}         & \multicolumn{1}{c|}{0.266}          & 0.094          & \multicolumn{1}{c|}{\textbf{102.264}} & \multicolumn{1}{c|}{\textbf{0.143}} & 0.166          & \multicolumn{1}{c|}{\textbf{11.426}} & \multicolumn{1}{c|}{0.387}          & 0.137          & \multicolumn{1}{c|}{\textbf{16.541}} & \multicolumn{1}{c|}{{\color[HTML]{000000} \textbf{0.193}}} & {\color[HTML]{000000} \textbf{0.123}} \\ \hline
\multicolumn{1}{|l|}{Diff-CSDI Uncond} & \multicolumn{1}{c|}{172.513}        & \multicolumn{1}{c|}{0.485}          & 0.137          & \multicolumn{1}{c|}{265.967}          & \multicolumn{1}{c|}{0.058}          & \textbf{0.138} & \multicolumn{1}{c|}{47.583}          & \multicolumn{1}{c|}{0.5}            & 0.212          & \multicolumn{1}{c|}{100.475}         & \multicolumn{1}{c|}{0.342}                                 & 0.178                                 \\ \hline
\end{tabular}}
\caption{Realism Comparison.}
\label{tab:realism}
\end{table*}

To quantitatively evaluate the realism of the generated volume distributions against empirical data, we employ three divergence metrics: Wasserstein distance, Kullback–Leibler (KL) divergence, and Kolmogorov–Smirnov (KS) statistic. The results, summarized in Table~\ref{tab:realism}, consistently indicate that the proposed diffusion-based models substantially outperform the GAN baseline in all four stocks, validating their effectiveness in capturing the empirical volume distribution. Specifically, \textit{Diff Cond} achieves slightly better performance than its unconditional counterpart \textit{Diff Uncond}, highlighting the benefit of incorporating conditioning information. Additionally, \textit{Diff Uncond} outperforms \textit{Diff-CSDI Uncond} model, a diffusion model variant based on the CSDI architecture \cite{tashiro2021csdi}, in almost all cases. This result underscores the superiority of our proposed DiffVolume architecture in modeling limit order book volume distributions. 

\subsection{Counterfactual Generation under Extreme Liquidity Scenarios}\label{subsection: counterfactual generation}

To evaluate the controllability of \textit{Diff Cond} under varying liquidity conditions, we design a counterfactual generation experiment based on liquidity-driven partitioning of the testing dataset. Specifically, we sort all snapshots into five bins (denoted as Bin 1 to Bin 5) based on ascending order of liquidity $\mathbf{\lambda}$. Bin 1 represents the lowest liquidity regime, whereas Bin 5 corresponds to the highest. The pairwise discrepancies are quantified using the Wasserstein distance, and the results are reported in Table~\ref{tab:counterfactual generation}.

We first compare \texttt{Real\_bin VS Fake\_bin}, where \texttt{Real\_bin} refers to the subset of real snapshots that fall into each of the five liquidity bins, and \texttt{Fake\_bin} denotes the corresponding generated samples from \textit{Diff Cond}, conditioned on the past volume history, time of day, and the actual future liquidity value associated with each snapshot in the bin. \texttt{Real\_all} denotes the entire testing dataset and \texttt{Fake\_all} denotes the generated dataset by following exact liquidity conditions of \texttt{Real\_all}. As shown in Table~\ref{tab:counterfactual generation}, the Wasserstein distances between \texttt{Real\_bin} and \texttt{Fake\_bin} are consistently lower than those between \texttt{Real\_all} and \texttt{Fake\_bin}. This confirms that \textit{Diff Cond} is highly effective in generating samples closely aligned with the intended liquidity regime.

To further assess the model’s ability to perform counterfactual generation, we fix the past volume trajectory $\mathbf{x}_{(t-L+1:t)}$ and time of day condition $\mathbf{\tau}$ while altering the target liquidity condition $\mathbf{\lambda}$ to simulate extreme scenarios:
\begin{itemize}[nosep,left=0pt]
    \item \texttt{Fake\_OL\_all} (Over Liquidity): For every snapshot in the test set, we randomly sample a future liquidity profile from Bin 5 (highest liquidity) and generate new snapshots using the fixed past conditions, time of day and sampled high liquidity.

\item \texttt{Fake\_UL\_all} (Under Liquidity): Similarly, we sample the liquidity condition from Bin 1 (lowest liquidity) while keeping the rest unchanged.
\end{itemize}
The results show clear trends. The Wasserstein distances between \texttt{Real\_bin} and \texttt{Fake\_OL\_all} decrease progressively from Bin 1 to Bin 5. This indicates that the samples generated under the fixed high liquidity condition resemble more closely those in the high-liquidity bin, validating the model’s ability to simulate high-liquidity scenarios regardless of the original snapshot conditions. Conversely, the distances between \texttt{Real\_bin} and \texttt{Fake\_UL\_all} increase from Bin 1 to Bin 5, suggesting that the low-liquidity condition effectively shifts the output distribution towards the characteristics of low-liquidity environments.

\smallskip \noindent \textbf{A Case Study on AAPL.} We observe that AAPL exhibits an anomalous pattern across bins 3 to 5. Specifically, the Wasserstein distances in \texttt{Real\_bin VS Fake\_bin} are unexpectedly higher than those in \texttt{Real\_all VS Fake\_bin}, which is contrary to the trend observed in MU, ADBE, and ZM. Furthermore, in the counterfactual setting (\texttt{Real\_bin VS Fake\_OL\_all}), the expected monotonic decrease in distance from bin 1 to bin 5 is not strictly preserved. Upon examining the underlying data distributions, we find that the liquidity profiles between AAPL’s training and testing datasets differ more substantially than in other stocks. In particular, the average liquidity values across the five bins in the training dataset are 
$\sim[1.08,1.36,1.54,1.73,2.07] \times 10^4$, while in the testing dataset they are $\sim[1.54,1.74,1.87,2.00,2.20]\times 10^4$. Notably, the average liquidity value of bin 3 in the training dataset is approximately equal to that of bin 1 in the testing dataset. This upward shift indicates a distributional drift in liquidity levels between the training and testing phases, which causes the model to misinterpret the intended conditioning during generation when extrapolating to high-liquidity regimes. 


\subsection{A Downstream task: Future Liquidity Prediction}\label{subsection: a downstream task}
To evaluate the utility of synthetic data generated by the proposed \textit{Diff Cond} model, we conduct a downstream task of short-term liquidity forecasting across the four stocks. Specifically, we aim to predict the aggregate LOB volume over the next $Horizon \in \{10, 30\}$ seconds, given a history window of 32 snapshots. Formally, the prediction target at time $t$ is defined as follows:
\begin{equation}
    y_t = \sum_{i=1}^{H} \sum_{j=1}^{20} x_{(t+i)}^{j}.
    \label{eq:downstream task target}
\end{equation}
We employ a LightGBM regressor trained on 27–28 February 2023, and test it on a held-out dataset on March 1, 2023. We compare two training scenarios: \texttt{Real} represents that the model is trained solely on the real dataset, while \texttt{Real + Synthetic} represents that the model is trained on an augmented dataset combining \texttt{Real\_all}, \texttt{Fake\_all}, \texttt{Fake\_OL\_all} and \texttt{Fake\_UL\_all} introduced in Subsection \ref{subsection: counterfactual generation}. Results are reported in Table~\ref{tab:liquidity prediction}, using Mean Squared Error (MSE), Mean Absolute Error (MAE), $R^2$, along with corresponding percentage improvement. By augmenting the training data with generated samples, we achieve consistent improvements in forecasting short-term liquidity. These results demonstrate that DiffVolume can also be practically beneficial by providing unseen samples in downstream prediction tasks.
\begin{table}\centering
\scalebox{0.85}{
\begin{tabular}{l|ccccc|}
\cline{2-6}
                                                  & \multicolumn{5}{c|}{Liquidity Quintile}                                                                                                                                                                  \\ \cline{2-6} 
                                                  & \multicolumn{1}{c|}{1}       & \multicolumn{1}{c|}{2}       & \multicolumn{1}{c|}{3}                              & \multicolumn{1}{c|}{4}                              & 5                              \\ \hline
\multicolumn{1}{|l|}{MU}                          &                              &                              &                                                     &                                                     &                                \\ \hline
\multicolumn{1}{|l|}{Real\_bin VS Fake\_bin}      & \multicolumn{1}{c|}{41.195}  & \multicolumn{1}{c|}{43.223}  & \multicolumn{1}{c|}{48.12}                          & \multicolumn{1}{c|}{46.884}                         & 54.666                         \\ \hline
\multicolumn{1}{|l|}{Real\_all VS Fake\_bin}      & \multicolumn{1}{c|}{163.107} & \multicolumn{1}{c|}{79.411}  & \multicolumn{1}{c|}{55.664}                         & \multicolumn{1}{c|}{108.858}                        & 154.762                        \\ \hline
\multicolumn{1}{|l|}{Real\_bin VS Fake\_OL\_all} & \multicolumn{1}{c|}{306.239} & \multicolumn{1}{c|}{215.215} & \multicolumn{1}{c|}{143.541}                        & \multicolumn{1}{c|}{142.479}                        & 137.262                        \\ \hline
\multicolumn{1}{|l|}{Real\_bin VS Fake\_UL\_all} & \multicolumn{1}{c|}{120.942} & \multicolumn{1}{c|}{206.558} & \multicolumn{1}{c|}{293.692}                        & \multicolumn{1}{c|}{357.38}                         & 426.005                        \\ \hline
\multicolumn{1}{|l|}{AAPL}                        &                              &                              &                                                     &                                                     &                                \\ \hline
\multicolumn{1}{|l|}{Real\_bin VS Fake\_bin}      & \multicolumn{1}{c|}{74.576}  & \multicolumn{1}{c|}{89.737}  & \multicolumn{1}{c|}{{\color[HTML]{000000} 105.091}} & \multicolumn{1}{c|}{{\color[HTML]{000000} 118.66}}  & {\color[HTML]{000000} 143.808} \\ \hline
\multicolumn{1}{|l|}{Real\_all VS Fake\_bin}      & \multicolumn{1}{c|}{223.924} & \multicolumn{1}{c|}{143.918} & \multicolumn{1}{c|}{{\color[HTML]{000000} 98.399}}  & \multicolumn{1}{c|}{{\color[HTML]{000000} 62.155}}  & {\color[HTML]{000000} 70.803}  \\ \hline
\multicolumn{1}{|l|}{Real\_bin VS Fake\_OL\_all} & \multicolumn{1}{c|}{194.573} & \multicolumn{1}{c|}{123.719} & \multicolumn{1}{c|}{{\color[HTML]{000000} 115.833}} & \multicolumn{1}{c|}{{\color[HTML]{000000} 134.759}} & {\color[HTML]{000000} 177.102} \\ \hline
\multicolumn{1}{|l|}{Real\_bin VS Fake\_UL\_all} & \multicolumn{1}{c|}{81.01}   & \multicolumn{1}{c|}{162.991} & \multicolumn{1}{c|}{225.618}                        & \multicolumn{1}{c|}{288.545}                        & 387.232                        \\ \hline
\multicolumn{1}{|l|}{ADBE}                        &                              &                              &                                                     &                                                     &                                \\ \hline
\multicolumn{1}{|l|}{Real\_bin VS Fake\_bin}      & \multicolumn{1}{c|}{9.566}   & \multicolumn{1}{c|}{10.172}  & \multicolumn{1}{c|}{10.817}                         & \multicolumn{1}{c|}{12.085}                         & 16.13                          \\ \hline
\multicolumn{1}{|l|}{Real\_all VS Fake\_bin}      & \multicolumn{1}{c|}{16.363}  & \multicolumn{1}{c|}{10.146}  & \multicolumn{1}{c|}{11.207}                         & \multicolumn{1}{c|}{15.574}                         & 26.742                         \\ \hline
\multicolumn{1}{|l|}{Real\_bin VS Fake\_OL\_all} & \multicolumn{1}{c|}{42.698}  & \multicolumn{1}{c|}{31.313}  & \multicolumn{1}{c|}{24.686}                         & \multicolumn{1}{c|}{18.839}                         & 14.745                         \\ \hline
\multicolumn{1}{|l|}{Real\_bin VS Fake\_UL\_all} & \multicolumn{1}{c|}{11.187}  & \multicolumn{1}{c|}{8.188}   & \multicolumn{1}{c|}{13.509}                         & \multicolumn{1}{c|}{21.307}                         & 36.276                         \\ \hline
\multicolumn{1}{|l|}{ZM}                          &                              &                              &                                                     &                                                     &                                \\ \hline
\multicolumn{1}{|l|}{Real\_bin VS Fake\_bin}      & \multicolumn{1}{c|}{13.958}  & \multicolumn{1}{c|}{15.316}  & \multicolumn{1}{c|}{15.024}                         & \multicolumn{1}{c|}{17.637}                         & 26.046                         \\ \hline
\multicolumn{1}{|l|}{Real\_all VS Fake\_bin}      & \multicolumn{1}{c|}{46.654}  & \multicolumn{1}{c|}{20.906}  & \multicolumn{1}{c|}{19.718}                         & \multicolumn{1}{c|}{28.967}                         & 65.739                         \\ \hline
\multicolumn{1}{|l|}{Real\_bin VS Fake\_OL\_all} & \multicolumn{1}{c|}{112.867} & \multicolumn{1}{c|}{84.289}  & \multicolumn{1}{c|}{67.306}                         & \multicolumn{1}{c|}{49.439}                         & 27.286                         \\ \hline
\multicolumn{1}{|l|}{Real\_bin VS Fake\_UL\_all} & \multicolumn{1}{c|}{16.607}  & \multicolumn{1}{c|}{26.199}  & \multicolumn{1}{c|}{43.166}                         & \multicolumn{1}{c|}{63.149}                         & 102.469                        \\ \hline
\end{tabular}}
\caption{Evaluation of Counterfactual Generation}
\label{tab:counterfactual generation}
\end{table}
\begin{table}\centering
\scalebox{0.85}{
\begin{tabular}{l|cccccc|}
\cline{2-7}
                                 & \multicolumn{3}{c|}{Horizon = 10}                                                                        & \multicolumn{3}{c|}{Horizon = 30}                                                     \\ \cline{2-7} 
                                 & \multicolumn{1}{c|}{MSE}       & \multicolumn{1}{c|}{MAE}    & \multicolumn{1}{c|}{$R^2$} & \multicolumn{1}{c|}{MSE}        & \multicolumn{1}{c|}{MAE}     & $R^2$ \\ \hline
\multicolumn{1}{|l|}{MU}         &                                &                             &                                           &                                 &                              &                      \\ \hline
\multicolumn{1}{|l|}{Real}       & \multicolumn{1}{c|}{162896410} & \multicolumn{1}{c|}{9841}   & \multicolumn{1}{c|}{0.7996}               & \multicolumn{1}{c|}{1584188284} & \multicolumn{1}{c|}{30962}   & 0.7782               \\ \hline
\multicolumn{1}{|l|}{Real + Synthetic}        & \multicolumn{1}{c|}{145313593} & \multicolumn{1}{c|}{9590}   & \multicolumn{1}{c|}{0.8212}               & \multicolumn{1}{c|}{1439704748} & \multicolumn{1}{c|}{30288}   & 0.7985               \\ \hline
\multicolumn{1}{|l|}{Percentage} & \multicolumn{1}{c|}{10.79\%}   & \multicolumn{1}{c|}{2.55\%} & \multicolumn{1}{c|}{2.70\%}               & \multicolumn{1}{c|}{9.12\%}     & \multicolumn{1}{c|}{2.17\%}  & 2.61\%               \\ \hline
\multicolumn{1}{|l|}{AAPL}       &                                &                             &                                           &                                 &                              &                      \\ \hline
\multicolumn{1}{|l|}{Real}       & \multicolumn{1}{c|}{180833992} & \multicolumn{1}{c|}{9831}   & \multicolumn{1}{c|}{0.717}                & \multicolumn{1}{c|}{1796092385} & \multicolumn{1}{c|}{31156}   & 0.6644               \\ \hline
\multicolumn{1}{|l|}{Real + Synthetic}        & \multicolumn{1}{c|}{137597830} & \multicolumn{1}{c|}{8892}   & \multicolumn{1}{c|}{0.7846}               & \multicolumn{1}{c|}{1601342963} & \multicolumn{1}{c|}{29751}   & 0.7008               \\ \hline
\multicolumn{1}{|l|}{Percentage} & \multicolumn{1}{c|}{23.91\%}   & \multicolumn{1}{c|}{9.55\%} & \multicolumn{1}{c|}{9.43\%}               & \multicolumn{1}{c|}{10.84\%}    & \multicolumn{1}{c|}{4.51\%}  & 5.48\%               \\ \hline
\multicolumn{1}{|l|}{ADBE}       &                                &                             &                                           &                                 &                              &                      \\ \hline
\multicolumn{1}{|l|}{Real}       & \multicolumn{1}{c|}{3358510}   & \multicolumn{1}{c|}{1464}   & \multicolumn{1}{c|}{0.4009}               & \multicolumn{1}{c|}{30205360}   & \multicolumn{1}{c|}{4422}    & 0.1906               \\ \hline
\multicolumn{1}{|l|}{Real + Synthetic}        & \multicolumn{1}{c|}{3239457}   & \multicolumn{1}{c|}{1428}   & \multicolumn{1}{c|}{0.4222}               & \multicolumn{1}{c|}{28691234}   & \multicolumn{1}{c|}{4228}    & 0.2312               \\ \hline
\multicolumn{1}{|l|}{Percentage} & \multicolumn{1}{c|}{3.54\%}    & \multicolumn{1}{c|}{2.46\%} & \multicolumn{1}{c|}{5.31\%}               & \multicolumn{1}{c|}{5.01\%}     & \multicolumn{1}{c|}{4.39\%}  & 21.30\%              \\ \hline
\multicolumn{1}{|l|}{ZM}         &                                &                             &                                           &                                 &                              &                      \\ \hline
\multicolumn{1}{|l|}{Real}       & \multicolumn{1}{c|}{705321941} & \multicolumn{1}{c|}{23866}  & \multicolumn{1}{c|}{-2.8648}              & \multicolumn{1}{c|}{7676911350} & \multicolumn{1}{c|}{81013}   & -4.3042              \\ \hline
\multicolumn{1}{|l|}{Real + Synthetic}        & \multicolumn{1}{c|}{652072041} & \multicolumn{1}{c|}{22684}  & \multicolumn{1}{c|}{-2.573}               & \multicolumn{1}{c|}{6195517329} & \multicolumn{1}{c|}{71388}   & -3.2806              \\ \hline
\multicolumn{1}{|l|}{Percentage} & \multicolumn{1}{c|}{7.55\%}    & \multicolumn{1}{c|}{4.95\%} & \multicolumn{1}{c|}{10.19\%}              & \multicolumn{1}{c|}{19.30\%}    & \multicolumn{1}{c|}{11.88\%} & 23.78\%              \\ \hline
\end{tabular}}
\caption{Liquidity Prediction Results.}
\label{tab:liquidity prediction}
\end{table}

\section{Conclusion}

In this paper, we propose a novel conditional denoising diffusion probabilistic model for generating future limit order book (LOB) volume snapshots, conditioned on historical order flow, time of day, and optionally target future liquidity levels. Through extensive empirical evaluations, we show that it achieves high realism, closely replicating statistical properties of real market volumes. Second, it supports counterfactual generation, enabling scenario-based simulations under hypothetical liquidity conditions. Third, it proves to be useful for a short-term liquidity prediction task by augmenting real data with synthetic samples. These results highlight the potential of conditional diffusion models as a robust and flexible generative framework for high-frequency financial modeling. Future work includes extending the model to multi-asset settings, joint price-volume generation, and applications to market simulation.

\section*{Acknowledgements}

The authors would like to thank Nikolas Nüsken from the Department of Mathematics at King’s College London for his insightful feedback and helpful suggestions on the manuscript.

\bibliographystyle{unsrtnat}  
\bibliography{references}

\end{document}